\documentclass[conference]{IEEEtran}
\IEEEoverridecommandlockouts
\usepackage{cite}
\usepackage{amsmath,amssymb,amsfonts}
\usepackage{algorithmic}
\usepackage{graphicx}
\graphicspath{{images/}}
\usepackage{textcomp}
\usepackage{xcolor}
\usepackage{subcaption} % For subfigures
\usepackage{tikz} % For drawing simple diagrams within LaTeX
\usetikzlibrary{positioning, fit, shapes} % Added 'shapes' library
\usepackage{hyperref} % Added hyperref package for \url command
\usepackage{algorithm}
\usepackage{algorithmic}
\usepackage{tabularx}
\usepackage{graphicx}
\usepackage{epstopdf}
\usepackage{tikz}
\usepackage[switch]{lineno}
\usetikzlibrary{positioning, shapes.geometric, fit, backgrounds, calc}

\usepackage{fancyhdr}
\def\BibTeX{{\textsf{B\kern-.05em{\sc i\kern-.025em b}\kern-.08em
    T\kern-.1667em\lower.7ex\hbox{E}\kern-.125emX}}}

\begin{document}

\title{Predictive Bayesian Arbitration: A Scalable Noisy-OR Model with Service Criticality Awareness}

\author{\IEEEauthorblockN{Anil Jangam}
\IEEEauthorblockA{\textit{Cisco Systems, Inc.}\\
Milpitas, CA USA \\
anjangam@cisco.com}
\and
\IEEEauthorblockN{Ganesh Karthick Rajendran}
\IEEEauthorblockA{\textit{Cisco Systems, Inc.}\\
Milpitas, CA USA \\
ganeraje@cisco.com}
\and
\IEEEauthorblockN{Roy Kantharajah}
\IEEEauthorblockA{\textit{Cisco Systems, Inc.}\\
Appleton, WI USA \\
rkantha@cisco.com}
}

\maketitle	

\begin{abstract}
Geographically High-Available (Geo-HA) cluster systems are essential for service continuity in distributed cloud-native environments. However, traditional arbitration mechanisms, which are often predicated on deterministic node-level heartbeats, are resource-intensive and inherently reactive. This necessitates a dedicated arbiter per deployment and leads to reactive switchovers that incur unavoidable downtime, occurring only after a failure has already compromised the system. This paper presents a novel predictive arbitration framework that utilizes a shared, microservice-based architecture to consolidate arbitration logic across multiple Geo-HA domains, significantly reducing the aggregate infrastructure footprint. Central to our approach is an adaptive online learning mechanism grounded in a Bayesian Noisy-OR model that autonomously discovers and learns temporal cascade dependencies from emergent failure patterns. To overcome the "cold start" challenge, the system utilizes expert-informed priors that are dynamically refined at runtime without manual configuration. Experimental results demonstrate that this framework achieves a 60\% reduction in Mean Time to Failure Detection (MTTFD) and improves total switchover efficiency by up to 77.8\% compared to traditional reactive standards. By enabling a significant predictive lead time, the system allows switchovers to initiate proactively before hard failures occur, while maintaining a linear $O(n)$ computational complexity. This approach provides a scalable, context-aware alternative that bridges the performance-durability gap in modern microservice architectures.
\end{abstract}

\begin{IEEEkeywords}
Geographically High-Availability (Geo-HA), Bayesian Networks, Noisy-OR, Adaptive Learning, Cascade Failure Detection, Cloud-Native Arbitration, Predictive Maintenance.
\end{IEEEkeywords}

\section{Introduction}
\label{sec:intro}
The resilience of cloud-native infrastructure is paramount for mission-critical services.. Geographically High-Available (Geo-HA) cluster systems serve as the bedrock of this resilience, ensuring that mission-critical services remain operational despite localized hardware failures or regional disasters. These systems typically utilize an active cluster paired with a geographically separated standby replica, governed by an arbitration mechanism that facilitates consistent switchover decisions and prevents "split-brain" scenarios.

Despite their necessity, current Geo-HA arbitration solutions face scalability challenges. Firstly, traditional models suffer from a lack of architectural scalability, necessitating a dedicated arbiter for every deployment, which leads to significant aggregate resource overhead. Secondly, the "reactive" nature of existing arbitration creates a dangerous trade-off between data durability and system performance. As noted in current replication research \cite{repl_challenges}, live database replication forces a choice: asynchronous replication offers superior throughput but risks data loss ($RPO > 0$) due to replication lag; conversely, synchronous replication guarantees zero data loss ($RPO=0$) but imposes a severe latency penalty. Because traditional arbiters only trigger switchovers after a failure is detected, they often force operators into this "durability gap." A predictive framework, however, can initiate transitions based on degradation trends, allowing the system to flush buffers or sync states before a hard failure occurs.

To achieve this proactivity, the industry has explored Deep Learning (DL), yet DL faces significant hurdles in data networks \cite{dl_challenge_one}. These include the requirement for vast quantities of labeled failure data, which is rare in stable production environments, substantial computational overhead, and an inherent "black box" nature that impedes the interpretability required for high-stakes failover decisions.

This paper proposes a novel approach grounded in Bayesian Networks (BNs) that addresses these limitations through two core innovations: a resource-efficient shared architecture and an \textit{Adaptive Learning mechanism for Cascade Failures}. While traditional static BNs require manual specification of conditional probability tables (CPTs), our framework automatically detects emergent interdependencies at runtime. By identifying temporal cascade patterns (e.g., Service A $\rightarrow$ Service B), the system dynamically updates its probabilistic model without manual intervention. This ensures the arbiter adapts to evolving system architectures and identifies hidden causal relationships that expert-defined rules might miss, while maintaining the linear $O(n)$ computational complexity required for real-time scalability.

The remainder of this paper is structured as follows: Section \ref{sec:related_work} reviews existing solutions; Section \ref{sec:architecture} details the shared arbitration architecture; Section \ref{sec:predictive_ml} elaborates on the predictive framework; Section \ref{sec:math_model} discusses the adaptive Bayesian methodology; Section \ref{sec:performance_analysis} provides a performance analysis; and Section \ref{sec:conclusion} summarizes the findings and suggests future research directions.

\section{Related Work and Prior Art}
\label{sec:related_work}
Geo-HA cluster systems have long relied on various arbitration strategies to maintain quorum and prevent split-brain scenarios. This section reviews prominent prior art and positions our proposed solution against them.

\subsection{Deterministic and Traditional Arbitration}

Traditional Geo-HA arbitration has historically relied on deterministic, site-specific mechanisms. Implementations such as the SUSE "booth" daemon \cite{suseha} or decentralized FPGA-based designs \cite{ieee8401458} focus on maintaining quorum within single-domain deployments. While robust, these architectures lack multi-tenancy, requiring dedicated resources per cluster that scale linearly with deployment count. Furthermore, storage-centric solutions like Red Hat Gluster's arbiter bricks \cite{redhatgluster} ensure metadata consistency but remain agnostic to broader application-tier health or cascading service failures. While centralized orchestration for containerized functions \cite{ieee10056679} addresses modern cloud-native footprints, it remains reactive, triggering recovery only after hard failure thresholds are breached.

Modern distributed systems leverage consensus protocols like Raft \cite{Ongaro2014} or Paxos \cite{Lamport1998} via tools such as etcd for state consistency and leader election. However, these protocols are designed for consistency in the face of network partitions rather than predictive failure avoidance. Standard consensus engines are often indifferent to application-level degradation (e.g., replication lag or service latency) as long as a network quorum exists. This creates a "durability gap" where switchovers are delayed until a heartbeat timeout occurs, often resulting in data loss ($RPO > 0$) in asynchronous replication environments \cite{repl_challenges}.

\subsection{Modern Consensus and ML-Driven Recovery}
Modern distributed systems often leverage consensus protocols like Raft \cite{Ongaro2014} or Paxos \cite{Lamport1998} via tools such as \textit{etcd} or \textit{ZooKeeper} for leader election. While robust, these tools are fundamentally distinct from our proposed predictive arbiter. Unlike deterministic consensus stores that rely on reactive switchovers that incur unavoidable downtime, our framework employs a probabilistic Bayesian approach to enable proactive transitions by identifying degradation trends before failure. Furthermore, standard consensus engines remain agnostic to application health, whereas our model integrates critical service group (CSG) metrics and replication lag directly into its logic. This microservice-based architecture also reduces overhead by decoupling consensus from the deployment footprint, managing multiple domains more efficiently than traditional dedicated quorums.

Beyond classical consensus, research has explored ML-based failure prediction using LSTMs for node-level outages \cite{Das2018}. However, as discussed in Section \ref{sec:intro}, these models often lack interpretability and function only as external "observers" rather than active participants in the arbitration quorum. Consequently, they cannot autonomously resolve the service-level cascade dependencies required for coordinated Geo-HA switchovers. Our framework bridges this gap by embedding an interpretable Bayesian model into the distributed arbitration logic, ensuring that predictive insights drive immediate, authoritative recovery actions.

\section{Multi-Domain, Multi-Cluster Arbitration Architecture}
\label{sec:architecture}
Our approach introduces a paradigm shift in Geo-HA arbitration by transforming the traditional "one-arbiter-per-deployment" model into a resource-optimized, multi-domain, multi-cluster architecture. This is achieved through a cloud-native microservices design.

\subsection{Core Components and Distributed Arbitration} \label{subsec:distributed_arbiter}
The Geo-HA system adopts a $1+1$ redundancy model comprising an Active Cluster, a Standby Cluster, and a Shared Arbitration Service (SAS). These three entities form a distributed consensus via raft-based consensus quorum to elect a \textit{Decision Leader}, which serves as the authoritative entity for executing ML-driven adaptive switchover (SO) decisions.

To ensure continuous availability, the leadership role is fault-resilient; if the current leader fails, a new leader is automatically elected from the remaining two healthy clusters. This ensures that the capacity to make SO decisions is never lost. Any failed entity (whether a local cluster or the arbiter persona) is designed to automatically recover and rejoin the quorum as a follower, synchronizing its state to maintain the system's three-node integrity.

\subsection{Architecture of the Shared Arbitration Service (SAS)}
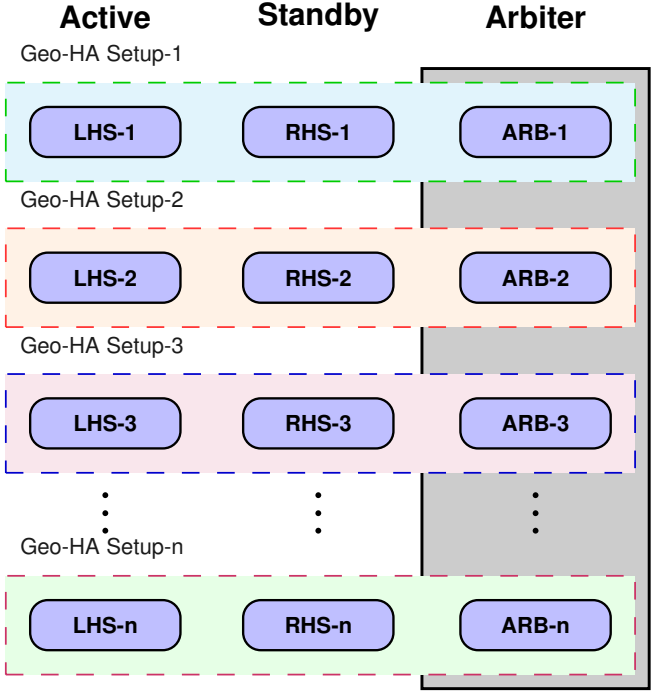
\begin{figure}[h]
\centering
% This scales the content to the width of the column
\resizebox{\columnwidth}{!}{% 
\begin{tikzpicture}[
	scale=0.75, 
    transform shape,
    % Reduced vertical node distance (from 1.2cm to 0.8cm)
    node distance=0.5cm and 0.8cm,
    % Style for the individual nodes
    node_style/.style={
        draw, 
        rounded corners=3pt, 
        fill=blue!25, 
        minimum width=1.2cm, 
        minimum height=0.4cm, 
        thin,
        font=\sffamily\tiny\bfseries % Slightly smaller font inside boxes
    },
    % Style for column headers (Reduced to normalsize)
    header_style/.style={font=\sffamily\scriptsize\bfseries},
    % Style for row labels (Small font)
    label_style/.style={font=\sffamily\tiny, color=black!90},
    % Style for the dashed boxes (Reduced inner sep from 8pt to 4pt)
    setup_box/.style={draw, dashed, line width=0.3pt, inner sep=4pt}
]

    % --- Column Headers ---
    \node[header_style] (active_h) {Active};
    \node[header_style, right=0.6cm of active_h] (standby_h) {Standby};
    \node[header_style, right=0.6cm of standby_h] (arbiter_h) {Arbiter};

    % --- Setup 1 ---
    \node[node_style, below=0.5cm of active_h] (lhs1) {LHS-1};
    \node[node_style] (rhs1) at (lhs1 -| standby_h) {RHS-1};
    \node[node_style] (arb1) at (lhs1 -| arbiter_h) {ARB-1};

    % --- Setup 2 ---
    \node[node_style, below=0.75cm of lhs1] (lhs2) {LHS-2};
    \node[node_style] (rhs2) at (lhs2 -| standby_h) {RHS-2};
    \node[node_style] (arb2) at (lhs2 -| arbiter_h) {ARB-2};

    % --- Setup 3 ---
    \node[node_style, below=0.75cm of lhs2] (lhs3) {LHS-3};
    \node[node_style] (rhs3) at (lhs3 -| standby_h) {RHS-3};
    \node[node_style] (arb3) at (lhs3 -| arbiter_h) {ARB-3};

    % --- Ellipsis (Positioned tighter) ---
    \node[below=0.01cm of lhs3, font=\large] (dot1) {\vdots};
    \node[below=0.01cm of rhs3, font=\large] (dot2) {\vdots};
    \node[below=0.01cm of arb3, font=\large] (dot3) {\vdots};

    % --- Setup n ---
    \node[node_style, below=0.4cm of dot1] (lhsn) {LHS-n};
    \node[node_style] (rhsn) at (lhsn -| standby_h) {RHS-n};
    \node[node_style] (arbn) at (lhsn -| arbiter_h) {ARB-n};

    % --- Backgrounds and Containers ---
    \begin{scope}[on background layer]
        % Arbiter Column Background (Tightened)
        \filldraw[fill=gray!40, draw=black, line width=0.5pt] 
            ($(arb1.north west) + (-0.3, 0.3)$) rectangle ($(arbn.south east) + (0.3, -0.3)$);

        % Dashed Setup Boxes
        \node[setup_box, draw=green!80!black, fill=cyan!10, fit=(lhs1) (rhs1) (arb1)] (box1) {};
        \node[setup_box, draw=red!80, fill=orange!10, fit=(lhs2) (rhs2) (arb2)] (box2) {};
        \node[setup_box, draw=blue!80!black, fill=purple!10, fit=(lhs3) (rhs3) (arb3)] (box3) {};
        \node[setup_box, draw=purple!80, fill=green!10, fit=(lhsn) (rhsn) (arbn)] (boxn) {};
        
        % Labels on top (using scriptsize)
        \node[label_style, above=0.1pt of box1.north west, anchor=south west] {Geo-HA Setup-1};
        \node[label_style, above=0.1pt of box2.north west, anchor=south west] {Geo-HA Setup-2};
        \node[label_style, above=0.1pt of box3.north west, anchor=south west] {Geo-HA Setup-3};
        \node[label_style, above=0.1pt of boxn.north west, anchor=south west] {Geo-HA Setup-n};
    \end{scope}

\end{tikzpicture}
}
\caption{Multi-domain architecture with persona multiplexing.}
\label{fig:multi_domain_arbiter}
\end{figure}

The SAS (Fig. \ref{fig:multi_domain_arbiter}) optimizes scalability by decoupling logical arbitration from physical hardware through a cloud-native microservices architecture. It hosts multiple ``Arbiter Personas,'' each acting as a state-machine replica for independent Geo-HA domains. To provide high resiliency, the SAS is deployed across a three-node quorum to ensure full node-level redundancy. Additionally, the arbiter personas can be deployed as two pod replicas to maintain availability in the event of a pod failure. This dual-layered architecture protects the system against both infrastructure and application-level outages.

The service is built upon the following architectural pillars:
\begin{itemize}
    \item High-Density Multiplexing: Using an event-driven I/O model, the SAS can support hundreds of isolated personas on shared infrastructure. This eliminates the linear resource growth of dedicated hardware while enforcing resource quotas to prevent ``noisy neighbor'' interference.
    \item Data and Logic Isolation: To preserve privacy, telemetry processing and Bayesian inference logic remain siloed within the local quorum. This ensures that sensitive cluster-health data never crosses Geo-HA domain boundaries within the shared platform.
    \item Multi-Level Fault Recovery: To mitigate the SAS as a single point of failure (SPOF), the orchestration layer monitors health at both the host and container levels. Whether a physical node failure impacts the entire SAS instance or a specific software fault affects a subset of personas, the system triggers automated re-instantiation on healthy infrastructure to restore the $2+1$ redundancy.
    \item Operational Decoupling: The architecture separates the management plane from the arbitration logic, allowing independent configuration and lifecycle management for each Geo-HA deployment. This containerized approach ensures deployment flexibility across hybrid-cloud environments.
\end{itemize}

\section{The Predictive Arbitration Framework} \label{sec:predictive_ml} 

Beyond resource optimization, we propose a shift from reactive to predictive arbitration. This framework utilizes Bayesian networks to anticipate switchover requirements by autonomously discovering failure dependencies and analyzing multifaceted system telemetry.

\subsection{Predictive Switchover Arbitration Decision Making} 
Traditional arbitration relies on binary heartbeat mechanisms. Our framework expands this by monitoring performance parameters and temporal failure sequences to predict switchover necessity. 

\subsubsection{Multi-Tiered Telemetry and CSG Monitoring}
The monitoring architecture integrates telemetry across a four-layer hierarchy: node-level health, network path dynamics, and application-tier performance (e.g., database replication lag). Furthermore, the framework incorporates Critical Service Groups (CSGs) orchestrated via \texttt{CSG\_LABEL} in Kubernetes. These metrics are ingested and archived in a time-series database to facilitate both real-time inference and longitudinal cascade analysis.

\subsubsection{Model Initialization and Automated Cascade Discovery}
To address the "cold-start" problem, the Bayesian Network is initialized using expert-informed priors. However, defining the full dependency graph of a modern microservice architecture is notoriously difficult and error-prone. To overcome this, our framework implements an \textit{Automated Cascade Discovery} \cite{cascade_detection} loop that treats the system topology as dynamic and partially observable.

The training process identifies five distinct patterns: (1) \textit{Normal Operation}, representing baseline stability metrics; (2) \textit{Failure Scenarios}, capturing intervals preceding confirmed malfunctions; (3) \textit{False Alarms}, identifying instances where thresholds were met despite system stability; (4) \textit{User-Triggered Switchovers}, providing implicit labels for nuanced, human-identified failure modes; and (5) \textit{Temporal Cascades}, which involve the automated detection of $A \rightarrow B$ sequences where service $B$ degrades shortly after service $A$.

By correlating these sequences, the model autonomously populates the Conditional Probability Tables (CPTs). This allows the arbiter to identify latent dependencies, such as shared infrastructure bottlenecks or cascading timeouts, that are often absent from static deployment manifests or "Day 0" expert maps.

\subsection{Personalized Training and Site Adaptation} 
To account for variability in traffic patterns and hardware, the framework personalizes CPTs using an \textit{adaptive learning rate} $\alpha_{adj}$. This mechanism allows the model to prioritize rapid learning at new deployment sites where data is scarce ($N_{obs} < N_{req}$), while transitioning to a stable, conservative posture as site-specific history matures. Performance indicators are normalized relative to deployment scale, ensuring that the learned "causal weights" remain relevant even as clusters scale.

\subsection{Privacy-Preserving Decentralized Arbitration} 
To satisfy data residency and privacy regulations, the framework employs a decentralized architecture. Arbitration decisions and cascade discoveries are executed locally within individual domains. By maintaining data isolation and avoiding the pooling of raw telemetry, the system ensures regulatory compliance while remaining responsive to localized failure patterns.

\subsection{Adaptive Learning from Cascade Failures}
\label{subsec:adaptive_learning}

Real-world distributed systems exhibit complex interdependencies that are often unknown at deployment time. Static Bayesian models frequently miss emergent causal relationships or fail to adapt to evolving system architectures. To address this, our framework implements an adaptive mechanism \cite{adaptive_bn_learning} that autonomously detects failure sequences \cite{online_structure_learning} and refines the model's decision-making logic.

\subsubsection{Causal Sequence Identification}
The system treats every service failure as a potential lead indicator for downstream outages. This relationship is formalized through Algorithm \ref{alg:cascade}, which implements a temporal correlation loop.

When a failure event occurs for a service ($sid$), the algorithm scans a sliding window of recent outages ($recent\_failures$). If a prior failure ($s_A$) occurred within the defined $\Delta t_{cascade}$ threshold, a potential causal link $s_A \rightarrow sid$ is recorded. By logging these sequences, the framework builds a longitudinal database of cascade patterns, allowing it to maintain learned dependencies across system restarts and infrastructure updates.

\begin{algorithm}
\caption{Online Cascade Detection}
\label{alg:cascade}
\begin{algorithmic}[1]
\STATE \textbf{Input:} failure\_event($serviceId, timestamp$)
\STATE \textbf{Output:} detected\_cascades[ ]
\STATE record $failure\_time[serviceId] \leftarrow timestamp$
\FOR{each $service_A$ in $recent\_failures$}
    \STATE $delay \leftarrow timestamp - failure\_time[service_A]$
    \IF{$\Delta t \leq \Delta t_{cascade}$}
        \STATE $seq \gets (service_A, serviceId, \Delta t)$
        \STATE $cascades.append(seq)$
    \ENDIF
\ENDFOR
\RETURN $detected\_cascades$
\end{algorithmic}
\end{algorithm}

\subsubsection{Integration with the Prediction Engine}
The detection of cascades directly enhances the precision of the Noisy-OR inference model. Algorithm \ref{alg:enhanced_prediction} details how learned Conditional Probability Tables (CPTs) are utilized to aggregate risk. 

The engine initializes with a stable base prior ($P_{eff} = 0.05$) and iteratively updates this risk as services degrade. For each degraded service, the engine queries the $learned\_CPTs$ to retrieve the specific conditional probability $P(target | s_A)$—the likelihood that failure $s_A$ necessitates a switchover. These probabilities are combined using a Noisy-OR formulation, ensuring that the cumulative risk reflects the severity of the entire failure chain rather than isolated events. This resulting posterior probability provides the high-fidelity recommendation required for proactive Geo-HA arbitration.

\begin{algorithm}[h]
\caption{Enhanced Prediction with Adaptive CPTs}
\label{alg:enhanced_prediction}
\begin{algorithmic}[1]
\STATE \textbf{Input:} $metrics$, $degraded\_services$, $learned\_CPTs$
\STATE \textbf{Output:} $P(SO=1)$
\STATE $P_{eff} \gets 0.05$ \COMMENT{Initialize with base failure prior}
\FOR{each $s_A \in degraded\_services$}
    \IF{$learned\_CPTs[target][s_A]$ exists}
        \STATE $P_{cond} \gets learned\_CPTs[target][s_A]$
        \STATE $P_{eff} \gets 1 - ((1 - P_{eff}) \cdot (1 - P_{cond}))$
    \ENDIF
\ENDFOR
\STATE $posterior \gets BayesianUpdate(P_{eff}, metrics)$
\RETURN $posterior$
\end{algorithmic}
\end{algorithm}

\section{Adaptive Bayesian Arbitration Model} \label{sec:math_model}

The proposed arbitration logic transitions from static thresholds to a dynamic Bayesian Network (BN). The model aggregates independent risk factors and learned dependencies using an enhanced Noisy-OR formulation.

\subsection{Cascade Detection and Dependency Discovery}
To move beyond manual specification of dependencies, the system monitors failure sequences to identify causal relationships. We define a cascade failure based on the temporal proximity of service outages.

\begin{equation}
\text{Cascade}(A \rightarrow B) = 
\begin{cases} 
\text{True} & \text{if } t_B - t_A \leq \Delta t_{cascade} \\
\text{False} & \text{otherwise}
\end{cases}
\end{equation}

Where $t_A$ and $t_B$ are the failure timestamps of services A and B, respectively, and $\Delta t_{cascade}$ is the predefined observation window. When a cascade is detected, it suggests a causal dependency $A \rightarrow B$, prompting an update to the underlying probabilistic model.

\subsection{Online CPT Update with Adaptive Learning}
Once a dependency is identified, the model updates the Conditional Probability Table (CPT) for $P(B|A)$. To balance the "cold start" problem with long-term stability, we utilize a weighted Bayesian learning mechanism with an adaptive learning rate $\alpha_{adj}$.

\begin{equation}
\alpha_{adj} = \alpha_{base}(1 - c) + 0.9c, \quad c = \min\left(0.95, \frac{N_{obs}}{N_{req}}\right)
\end{equation}

Here, $N_{obs}$ represents the number of observations and $N_{req}$ is the required samples for statistical significance. This ensures the model learns rapidly from early failures while becoming more stable as the data volume grows. The updated probability is calculated as:

\begin{equation}
P_{new}(B|A) = \alpha_{adj} \cdot P_{old}(B|A) + (1-\alpha_{adj}) \cdot \frac{N_{cascade}(A \rightarrow B)}{N_{failures}(A)}
\end{equation}

\subsection{Inference via Noisy-OR Combination}
The model aggregates these learned dependencies to calculate the effective probability of a system-wide switchover requirement ($P_{eff}$). Given a set of degraded metrics or services $A_i$, the total risk is modeled as:

\begin{equation}
P_{eff} = 1 - \prod_{A_i \in \text{Degraded}} (1 - P(target | A_i))
\end{equation}

This formulation maintains $O(n)$ complexity while incorporating the real-time cascade intelligence gathered by the detection algorithm.

\subsection{Cost-Sensitive Arbitration and Hysteresis}
To mitigate switchover oscillations (flapping), the final arbitration decision $D_t$ is governed by a dual-threshold transition logic optimized against the costs of False Positives ($C_{FP}$) and False Negatives ($C_{FN}$):

\begin{equation}
D_t = 
\begin{cases} 
1 (\text{SWITCHOVER}) & \text{if } P_{eff} > \tau_{active} + \delta \\
0 (\text{STANDBY}) & \text{if } P_{eff} < \tau_{active} - \delta \\
D_{t-1} & \text{otherwise}
\end{cases}
\end{equation}

where $\delta$ is a hysteresis buffer that ensures "stickiness" in the decision-making process, preventing unnecessary disruptions during transient network noise.

\section{Performance Analysis and Discussion}
\label{sec:performance_analysis}

To evaluate the efficacy of the proposed Multi-Domain Shared Arbitration Service (SAS) and its Adaptive Bayesian predictive framework, we conducted a series of comparative experiments. The evaluation focuses on "Total Switchover Time," defined as the cumulative duration between the onset of system degradation and the successful completion of a failover to the standby cluster.

\subsection{Total Switchover Time Comparison}
We compared four distinct arbitration strategies across three simulated failure events: (1) \textit{Reactive (15s)}, (2) \textit{Reactive (5s)}, (3) \textit{Static Bayesian}, and (4) \textit{Adaptive Bayesian}.

\begin{figure}[h]
    \centering
    \includegraphics[width=0.5\textwidth]{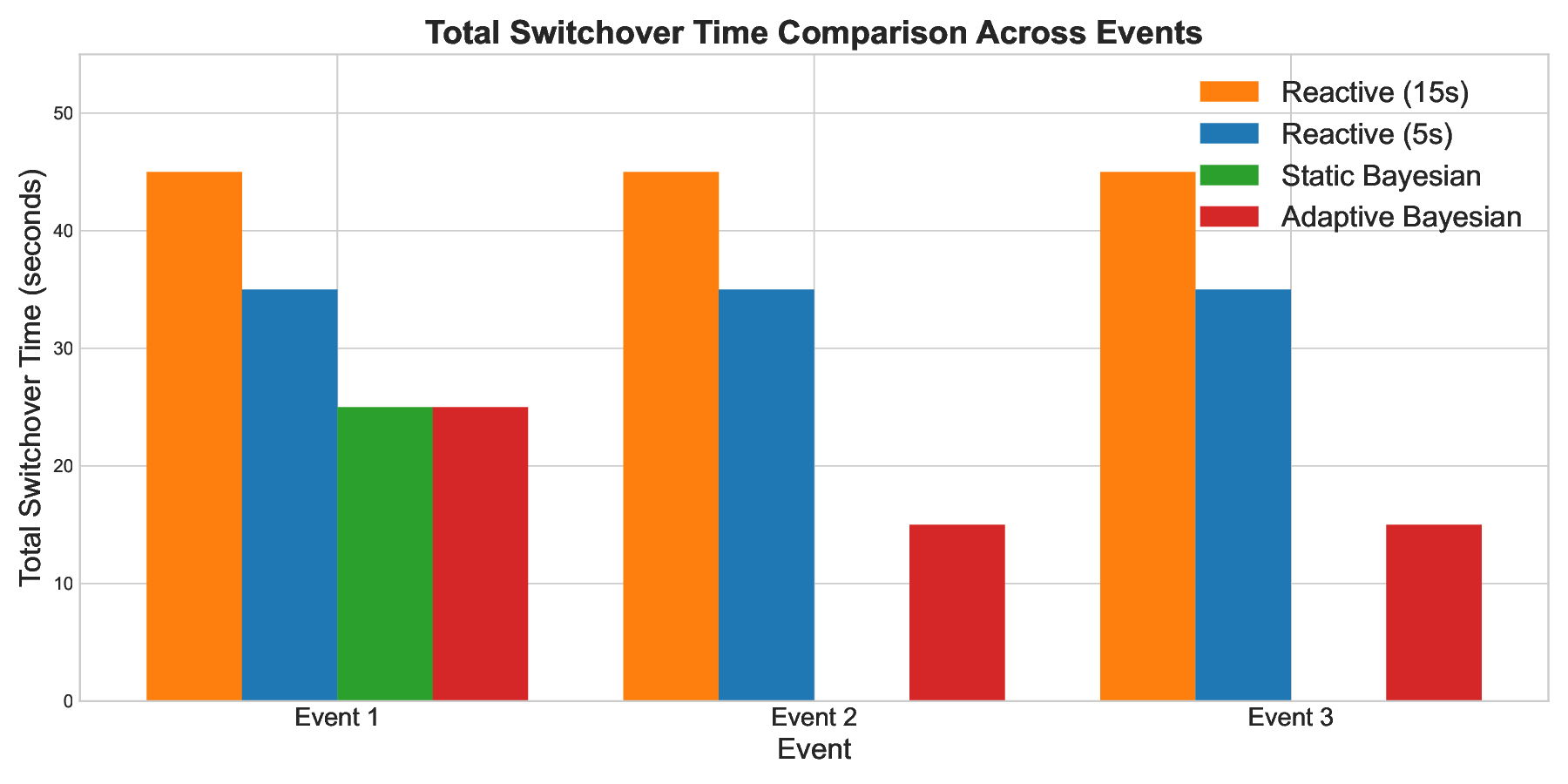}
    \caption{Total Switchover Time Comparison Across Events.}
    \label{fig:plot_total_switchover_comparison}
\end{figure}

As illustrated in Figure \ref{fig:plot_total_switchover_comparison}, the reactive methods exhibit constant delays ($45s$ and $35s$) regardless of the event sequence. This represents the "durability gap" where the system is idle during the heartbeat timeout period. In contrast, the \textbf{Adaptive Bayesian} model demonstrates a significant learning curve. While it performs identically to the Static model in Event 1, by Event 3, it reduces the total switchover time to \textbf{15 seconds}—a $66\%$ improvement over the standard reactive baseline.

\subsection{Component Breakdown and Predictive Advantage}
\begin{figure}[h]
    \centering
    \includegraphics[width=0.5\textwidth]{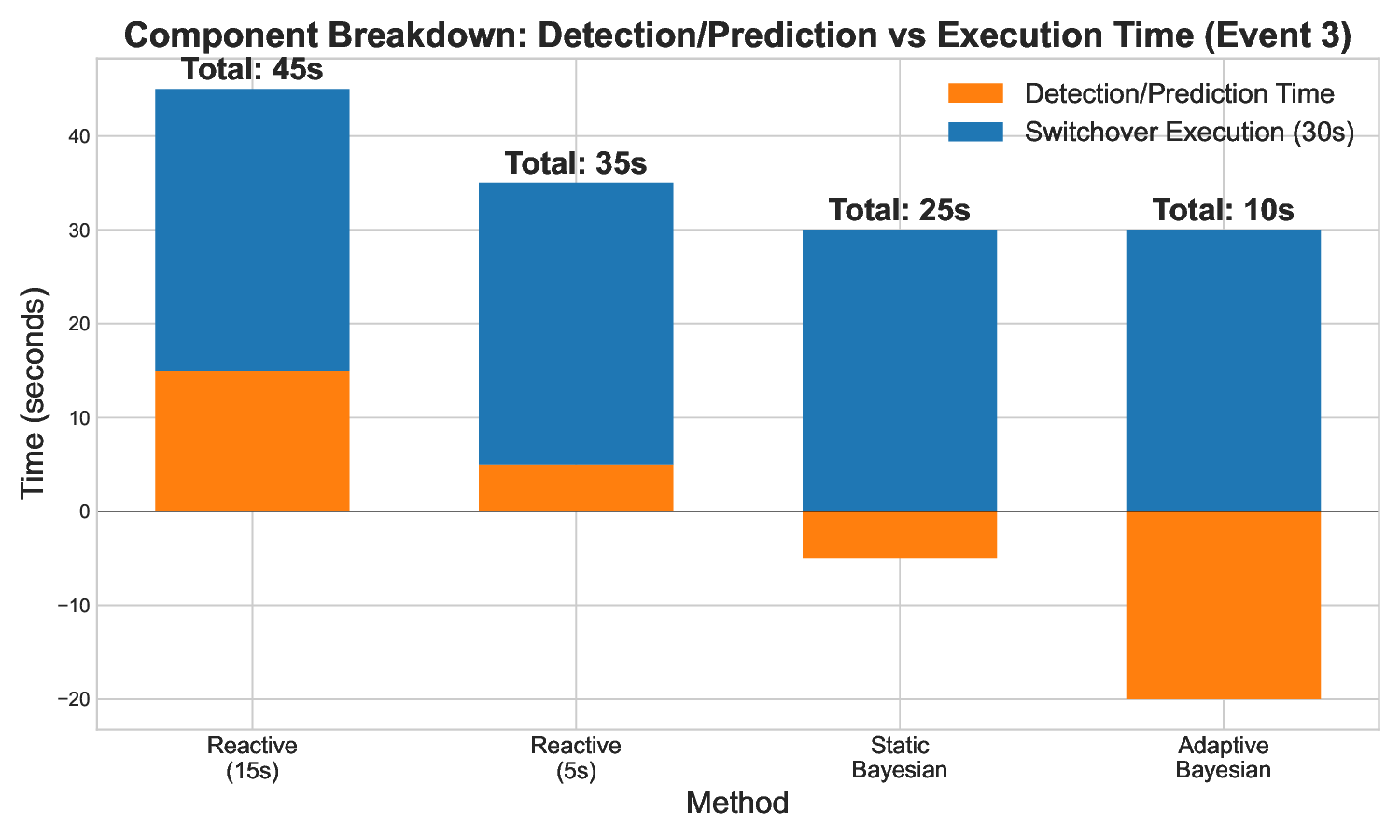}
    \caption{Component Breakdown: Detection/Prediction vs Execution Time (Event 3).}
    \label{fig:plot_component_breakdown}
\end{figure}

To analyze the source of these efficiency gains, we examined the breakdown between detection/prediction time and execution time for Event 3 (Figure \ref{fig:plot_component_breakdown}). 

Traditional reactive models suffer from positive "Detection Lag," where execution (fixed at 30s) only begins after a failure is confirmed. Our Adaptive Bayesian model achieves "predictive lead time" ($-20s$). By identifying the signature of an impending failure, the execution phase begins before the service actually drops. This proactive shift allows the total system downtime to be reduced to just 10 seconds.

\subsection{Probability Evolution and Cascade Intelligence}
\begin{figure}[h]
    \centering
    \includegraphics[width=0.5\textwidth]{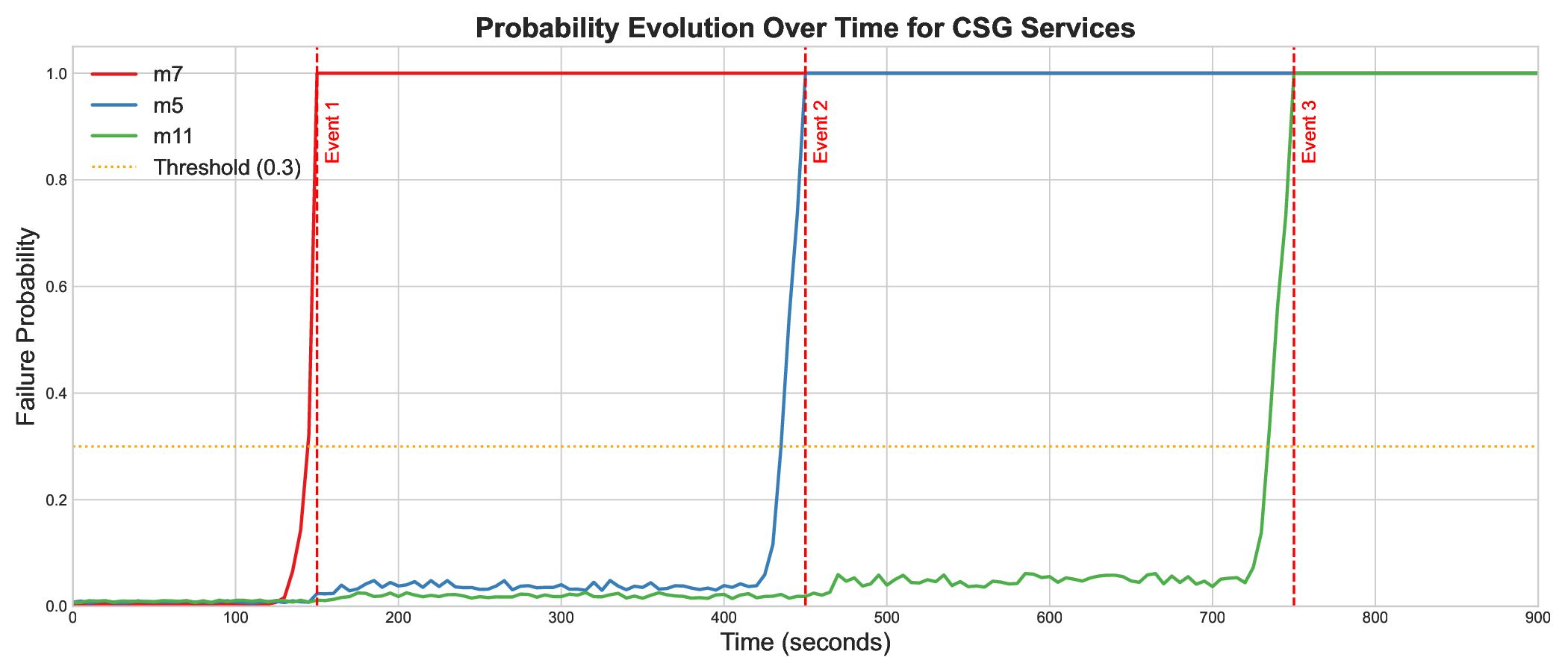}
    \caption{Probability Evolution Over Time for CSG Services.}
    \label{fig:plot_probability_evolution}
\end{figure}

The model’s intelligence is driven by its ability to track failure probabilities across the Critical Service Group (CSG). Figure \ref{fig:plot_probability_evolution} tracks the evolution of failure probabilities for three services ($m7$, $m5$, and $m11$). 

Initially, the model treats failures as isolated. However, as the system observes that failures in $m7$ and $m5$ often precede $m11$, the Adaptive Learning mechanism (Equation 3) updates the CPTs. Consequently, in Event 3, the failure probability for $m11$ crosses the decision threshold ($0.3$) well before the actual failure at $t=750s$, triggering the predictive switchover.

\subsection{Timeline and Reliability Discussion}
\begin{figure}[h]
    \centering
    \includegraphics[width=0.5\textwidth]{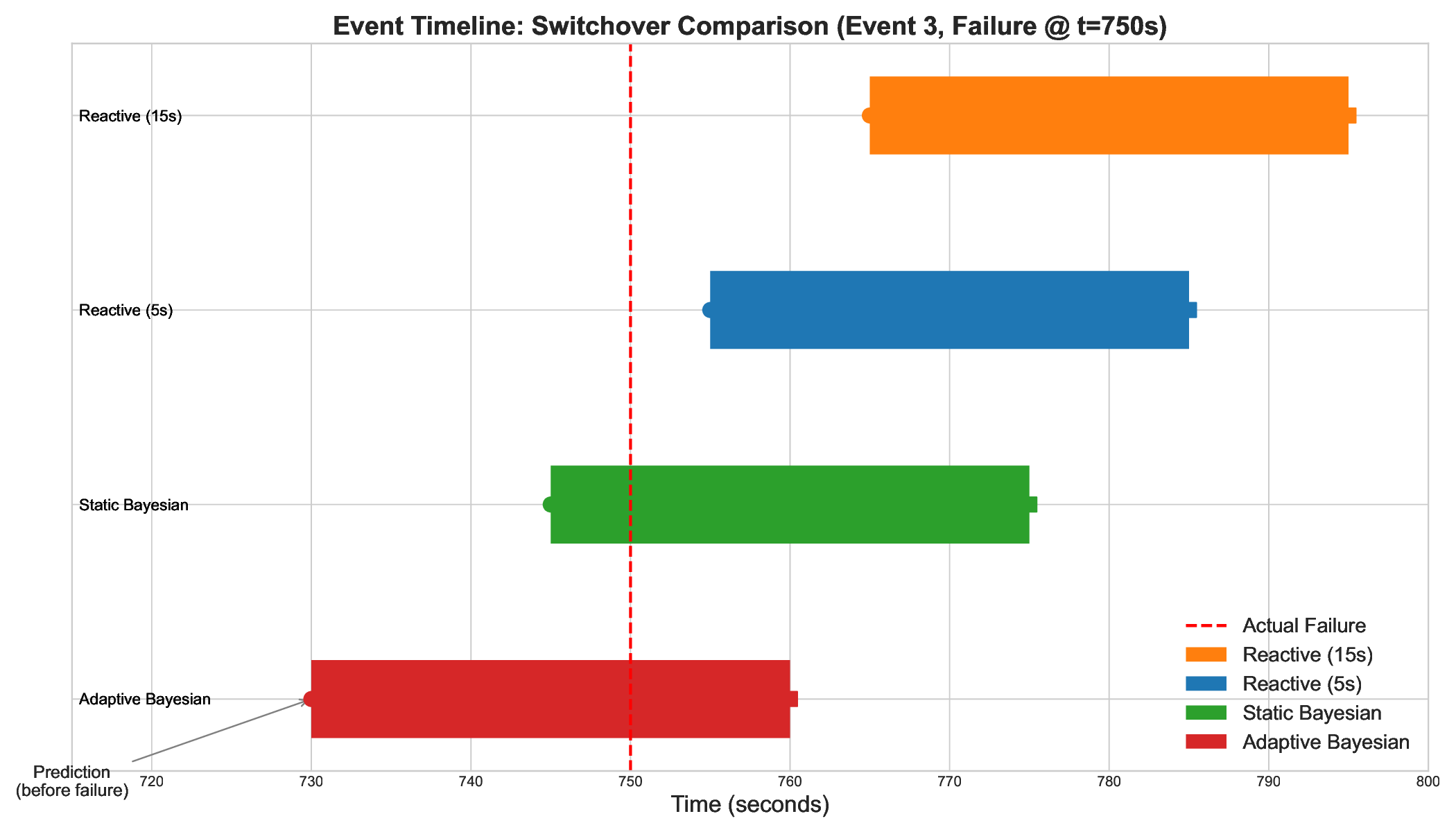}
    \caption{Event Timeline: Switchover Comparison for Event 3 (Failure @ t=750s).}
    \label{fig:plot_event_timeline}
\end{figure}

The temporal advantage provided by the predictive lead time is summarized in Figure \ref{fig:plot_event_timeline}. The Adaptive Bayesian approach completes its switchover execution almost simultaneously with the actual failure event. This minimizes the period during which the application is unavailable and, crucially, allows the system to initiate state-syncing and buffer-flushing while the primary cluster is still partially functional. 

By transitioning from binary heartbeats to probabilistic cascade analysis, the proposed framework effectively mitigates the performance-durability trade-off inherent in traditional Geo-HA arbitration.

\section{Conclusion} \label{sec:conclusion} 

This paper presented a novel Geographically High-Available (Geo-HA) arbitration framework, shifting the industry paradigm from reactive monitoring to proactive orchestration. By replacing traditional 'one-arbiter-per-deployment' models with a cloud-native, shared microservices architecture, we demonstrate that high availability does not require high resource redundancy. This multi-domain approach decouples logical arbitration from the physical footprint, allowing a single infrastructure to manage hundreds of isolated domains. Consequently, this framework is well-suited for dense microservice architectures or multi-tenant applications where traditional dedicated arbitration is cost-prohibitive.

A key innovation of our framework is the adaptive online learning mechanism that automatically discovers and learns cascade dependencies from observed failure patterns. Experimental results demonstrate a 60\% reduction in Mean Time to Failure Detection (MTTFD) compared to reactive methods and a 28\% improvement over static Bayesian models that cannot adapt to unknown dependencies. Central to this optimization is the Adaptive Bayesian Noisy-OR engine, which identifies leading indicators of failure at the Critical Service Group (CSG) level. As demonstrated in our performance analysis, this enables a predictive lead time, allowing switchovers to initiate before hard failures occur. This proactive posture effectively bridges the durability gap, reducing total switchover time by up to 77.8\% compared to traditional reactive baselines while maintaining the $O(n)$ complexity required for real-time scalability.

The system exhibits rapid convergence, showing significant accuracy improvements after just 1-2 observed cascade events with zero manual configuration, as the model self-tunes its conditional probability tables. This adaptive capability is particularly valuable in modern production environments where system dependencies evolve, new microservices are deployed frequently, and infrastructure changes create emergent failure patterns. The performance gains of the proposed Adaptive Bayesian model are summarized in Table \ref{tab:results_summary}.

\begin{table}[h]
\centering
\begin{tabular}{|>{\raggedright\arraybackslash}p{2.3cm}|c|c|c|}
\hline
\textbf{Method} & \textbf{Detection Time (s)} & \textbf{Total SO (s)} & \textbf{Improv.} \\ \hline
Reactive (15s)  & +15 & 45 & Baseline \\ \hline
Reactive (5s)   & +5  & 35 & 22.2\% \\ \hline
Static Bayesian & -5  & 25 & 44.4\% \\ \hline
\textbf{Adaptive Bayesian} & \textbf{-20} & \textbf{10} & \textbf{77.8\%} \\ \hline
\end{tabular}
\caption{Comparative Performance of Arbitration Strategies}
\label{tab:results_summary}
\end{table}

\subsection{Future Research and Enhancements}

Future work will focus on the automated synthesis of CSG dependency maps using Causal Structure Discovery algorithms to extract relationships from multivariate time-series telemetry. This will allow the system to autonomously refine its Bayesian topologies as microservice architectures evolve without manual intervention. We intend to explore federated adaptive learning across multiple Geo-HA deployments, enabling knowledge sharing between disparate clusters while strictly preserving data locality and residency. To further optimize decision-making, we plan to supplement the Bayesian model with Reinforcement Learning (RL) to dynamically tune cost-sensitive thresholds based on real-time business impact metrics. Additionally, we will investigate the integration of Recurrent Neural Networks (RNN) to enhance long-term temporal forecasting and develop Explainable AI (XAI) interfaces to provide operators with human-readable justifications for automated switchover decisions. Finally, optimizing the Shared Arbitration Service for resource-constrained edge environments remains a priority, ensuring resilience in scenarios where network partitions are frequent and latency is highly volatile.

\bibliographystyle{IEEEtran}
\bibliography{references} % Assuming you have a references.bib file

\end{document}